# Technical Report:
# Developing a Working Data Hub

Vijay Gadepally, Jeremy Kepner

{vijayg, kepner} @ ll.mit.edu

MIT Lincoln Laboratory Supercomputing Center

Lexington, MA 02421

March 2020

# EXECUTIVE SUMMARY

Data forms a key component of any enterprise. The need for high quality and easy access to data is further amplified by organizations wishing to leverage machine learning or artificial intelligence for their operations. To this end, many organizations are building resources for managing heterogenous data, providing end-users with an organization wide view of available data, and acting as a centralized repository for data owned/collected by an organization.

Very broadly, we refer to these class of techniques as a "data hub." While there is no clear definition of what constitutes a data hub, some of the key characteristics include:

- Data catalog
- Links to datasets or owners of data sets or centralized data repository
- Basic ability to serve / visualize data sets
- Access control policies that ensure secure data access and respects policies of data owners
- Computing capabilities tied with data hub infrastructure

Of course, developing such a data hub entails numerous challenges. This document provides background in databases, data management and outlines best practices and recommendations for developing and deploying a working data hub. A few key recommendations:

**Technology:**

- Support federated data access
- Support complex access control
- Tie computing and data hub
- Support multiple data formats

**Infrastructure:**

- Provision hardware and software correctly
- Leverage cloud computing judiciously
- Be wary of data sensitivity and different network security requirements

**Data Formatting:**

- Leverage human and machine-readable formats for data being shared
- Stick with simple conventions for file naming

**Security/Data Sharing:**

- Work with Information Security Officers from the beginning
- Integrate Subject Matter Expert feedback with ISO requirements
- Ensure data use agreements are in place when sharing with external parties

**Policy:**

- Create an acquisition environment conducive to new technology
- Prioritize open-source vs proprietary products
- Include management in technology selection
- Avoid software/hardware products without robust user and developer base

**User Engagement and Incentives:**

- Incorporate user feedback early in the development process and include key influencers
- Make data sharing/maintenance a key part of individual performance assessments
- Remove cost from the data sharing equation
- Leverage open, non-proprietary standards
- Invest in data discovery techniques
- Ask projects about data sharing plans during funding/proposal phase

Many organizations are investing in diverse tools to provide end-users with access to data. These data hub efforts can run into a number of challenges ranging from technology selection to user engagement. We believe that by incorporating the recommendations above with technology solutions outlined in this document, developers of data hub solutions can greatly increase the likelihood of success.

# TABLE OF CONTENTS

# 1. INTRODUCTION

The recent emergence of Artificial Intelligence as a field has been largely driven by the availability of large quantities of curated datasets [1-3], advanced algorithms such as neural networks [4], and computing hardware such as that available in high performance computing centers [5] or cloud infrastructure [6] [7] [8]. Working with Big Data is prone to a variety of challenges. Very often, these challenges are referred to as the three Vs of Big Data: Volume, Velocity and Variety. Most recently, there has been a new emergent challenge (perhaps a fourth V): Veracity [9]. These combined challenges constitute a large reason why Big Data is difficult to work with.

Big data volume stresses the storage, memory, and computational capacity of a computing system and often requires access to a computing cloud. The National Institute of Science and Technology (NIST) defines cloud computing to be "a model for enabling ubiquitous, convenient, on-demand network access to a shared pool of configurable computing resources ... that can be rapidly provisioned and released with minimal management effort or service provider interaction" [47]. Within this definition, there are different cloud models that satisfy different problem characteristics and choosing the right cloud model is problem specific. Currently, there are four multi-billion-dollar ecosystems that dominate the cloud-computing landscape: enterprise clouds, big data clouds, SQL database clouds, and supercomputing clouds. Each cloud ecosystem has its own hardware, software, conferences, and business markets. The broad nature of enterprise big data challenges make it unlikely that one cloud ecosystem can meet its needs, and solutions are likely to require the tools and techniques from more than one cloud ecosystem. For this reason, at the Massachusetts Institute of Technology (MIT) Lincoln Laboratory, we developed the MIT SuperCloud architecture [10-12] that enables the prototyping of four common computing ecosystems on a shared hardware platform. The velocity of big data stresses the rate at which data can be absorbed and meaningful answers produced. Very often, the velocity challenge is mitigated through high performance databases, file systems and/or processing. Big data variety may present the largest challenge and greatest opportunities. The promise of big data is the ability to correlate diverse and heterogeneous data to form new insights. A new fourth V [13], veracity challenges our ability to perform computation on these complex datasets while preserving trust in the data and analytic. These V's of big data can be amplified in applications such as autonomous vehicles [14], [15] and internet-of-things [16] enabled applications such as smart cities [17] [18] where quick decisions based on multiple sensor inputs are needed.

For an enterprise, these challenges can be mitigated by providing suitable infrastructure that allow users to collect, store and serve datasets of interest. As such, many organizations are currently attempting to provide a centralized resource that users can upload and download datasets of interest. These data lakes or data hubs aim to provide a one-stop solution for users looking to store or use datasets across an organization. Some publicly available examples include the Dataverse Project [19], the Kaggle competition website (https://www.kaggle.com/) and the US Government's *data.gov* [20]. Additionally, there are domain specific examples such as the Omics Discovery Index [21] and Nature's Scientific Data (https://www.nature.com/sdata/policies/repositories) catalog that links to a number of high quality datasets. A good overview of the state of dataset search is provided in [22].

A recent effort from Google highlights a number challenges that go in to developing a data catalog that include [23]:

- Maintaining quality: Different groups use different standards for labeling data and metadata.
- Duplication of results: There may be different versions of the same dataset available
- Data quality and provenance: Different groups have different quality standards
- Churn and Stale Sites
- Ranking and relevance
- Multiple metadata standards
- Security and sensitivity of data products
- Access control

The first step in developing a usable data hub involves decisions on storage and data management. Databases and filesystems provide access to vast amounts of data but differ at a fundamental level. Filesystem storage engines are designed to provide access to a potentially large subset of the full dataset. Database engines are designed to index and provide access to a smaller, but well defined, subset of data. Before looking at particular storage and database engines, it is important to take a look at where these systems fall within the larger big data system.

An example of an overall system architecture is given in Figure 1. At the lowest level, one needs the ability to bring together files of different types stored in hard drives, external data sources, databases, and data warehouses. Above these data sources is a data hub and platform later that provides heterogenous data management and rudimentary data transformations such that users can query multiple sources at once. The layer above that provide users with exploratory analytics, basic data integration capabilities and the ability to discover, link and clean datasets. It is important to note that all of these components are required to achieve the desired end-result of easy and efficient data access for end users.

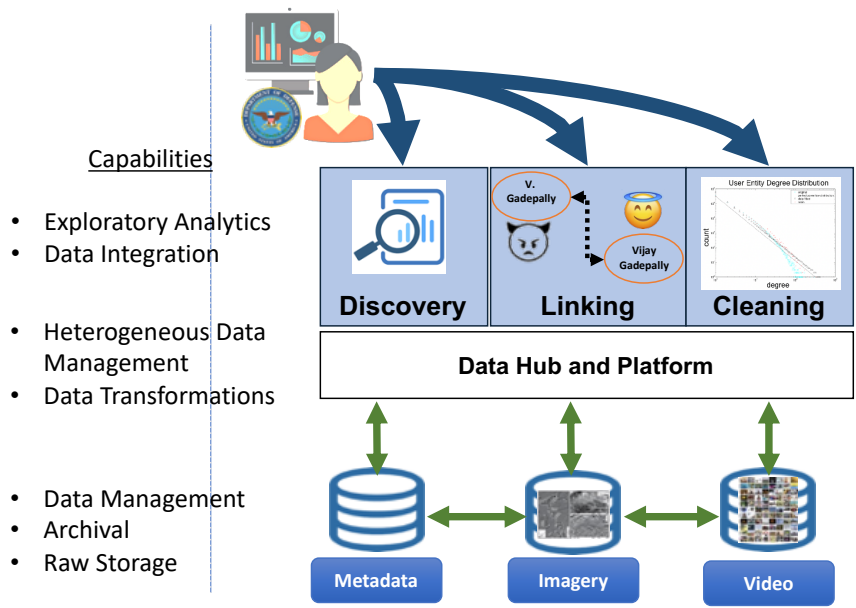

*Figure 1: High level view of data hub platform with various components*

This article focuses on the first two steps of this architecture and is outlined as follows. Section 0 provides a high-level system's engineering view of an end-to-end big data system. Section 3 provides an overview of filesystem-based storage. Section 4 outlines database management systems. Section 5 discusses trends in heterogenous data management systems. Finally, Section 6 outlines practical steps that can be taken to maximize the likelihood of success when developing a data hub.

## 2. SYSTEM ENGINEERING FOR DATA MANAGEMENT

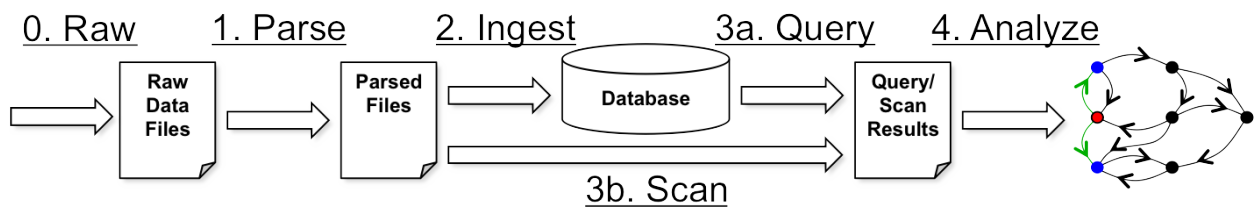

*Figure 2: A standard big data pipeline consists of five steps to go from raw data to useful analytics.*

Systems engineering studies the development of complex systems. Given the many challenges of Big Data as described in Section 1, systems engineering has a great deal of applicability to developing a Big Data system. Once convenient way to visualize a Big Data system is as a pipeline. In fact, most Big Data systems consist of different steps which are connected to each other to form a pipeline (sometimes, they may not be explicitly separated though that is the function they are performing). Figure 2 shows a notional pipeline for Big Data processing.

First, raw data is often collected from sensors or other such sources. These raw files often come in a variety of formats such as comma separated values (CSV), JavaScript Object Notation (JSON), or other proprietary sensor formats. Most often, this raw data is collected by the system and placed into files that replicate the formatting of the original sensor. Retrieval of raw data may be done by different interfaces such as cURL (http://curl.haxx.se/) or other messaging paradigms such as publish/subscribe. The aforementioned formats and retrieval interfaces are by no means exhaustive but highlight some of the popular tools being used.

Once the raw data is on the target system, the next step in the pipeline is to parse these files into a more readable format or to remove components that are not required for the end-analytic. Often, this step involves removing remnants of the original data collection step such as unique identifiers that are no longer needed for further processing. The parsed files are often kept on a serial or parallel file system and can be used directly for analytics by scanning files. For example, a simple word count analytic can be done by using the Linux grep command on the parsed files, or more complex analytics can be performed by using a parallel processing framework such as Hadoop MapReduce [24] or the Message Passing Interface (MPI) [25]. As an example of an analytic which works best directly with the file system, dimensional analysis [26] performs aggregate statistics on the full dataset and is much more efficient working directly from a high-performance parallel file system.

For other analytics (especially those that wish to access only a small portion of the entire dataset), it is convenient to ingest this data into a suitable database. An example of such an analytic is given in [27] which performs an analysis on the popularity of particular entities in a database. This example takes only a small, random piece of the dataset (the counts of words is much smaller than the full dataset) and is well suited for database usage. Once data is in the database or on the filesystem, a user can write queries or scans depending on their use case to produce results that can then be used for complex analytics such as topic modeling.

Each step of the pipeline involves a variety of choices and decisions. These choices may depend on hardware, software or other factors. Many of these choices will also make a difference to the later parts of the pipeline and it is important to make informed decisions. Some of the choices that one may have at each step:

- Step 0: Size of individual raw data files, output format
- Step 1: Parsed data contents, data representation, parser design
- Step 2: Size of database, number of parallel processors, pre-processing
- Step 3: Scan or query for data, use of parallel processing
- Step 4: Visualization tools, algorithms

# 3. RAW DATA STORAGE

One of the most common ways to store a large quantity of data is through the use of traditional storage media such as hard drives. There are many storage options that must be carefully considered that depend upon various parameters such as total data volume and desired read and write rates. In the pipeline of Figure 2, the storage engine plays an important part of steps two and three. A more detailed view of these techniques is provided in [28] and [29].

In order to deal with many challenges such as preserving data through failures, the past decades have seen the development of many technologies such as RAID (redundant array of independent disks) [30], NFS (network file system), HDFS (Hadoop Distributed File System) [31], and Lustre [32]. These technologies aim to abstract the physical hardware away from application developers in order to provide an interface for an operating system to keep track of a large number of files while allowing support for data failure, high speed seeks, and fast writes. As an exemplar technology, we will discuss the Lustre distributed file storage system.

## Serial Memory and Storage

The most prevalent form of data storage is provided by an individual's laptop or desktop system. Within these systems, there are different levels of memory and storage that trade off speed with cost calculated as bytes per dollar. The fastest memory provided by a system (apart from the relatively low capacity system cache) is the main memory or random access memory (RAM). This volatile memory provides relatively high speed (10s of GB/s in 2020) and is often used to store data up to terabytes in 2020. When the data size is larger than the main memory, other forms of storage are used. Within serial storage technologies, some of the most common are traditional spinning magnetic disc hard drives and solid state drives (solid state drives may be designed to use volatile RAM or non-volatile flash technology). The capacity of these technologies can be in the 100s of TB each and can support transfer rates anywhere from approximately 200MB/s to multiple GB/s in 2020.

## Distributed Storage Technology Example: Lustre

Lustre (http://lustre.org/) is a distributed filesystem technology designed to meet the highest bandwidth file requirements on the largest systems in the world and is used for a variety of scientific workloads [33]. The open source Lustre parallel file system presents itself as a standard POSIX, general-purpose file system and is mounted by client computers running the Lustre client software. Files stored in Lustre contain two components: metadata and object data. Metadata consists of the fields associated with each file such as i-node, filename, file permissions, and timestamps. Object data consists of the binary data stored in the file. File metadata is stored in the Lustre metadata server (MDS). Object data is stored in object storage servers (OSSes) shown in Figure 3. When a client requests data from a file, it first contacts the MDS which returns pointers to the appropriate objects in the OSSes. This movement of information is transparent to the user and handled fully by the Lustre client. To an application, Lustre operations appear as standard file system operations and require no modification of application code.

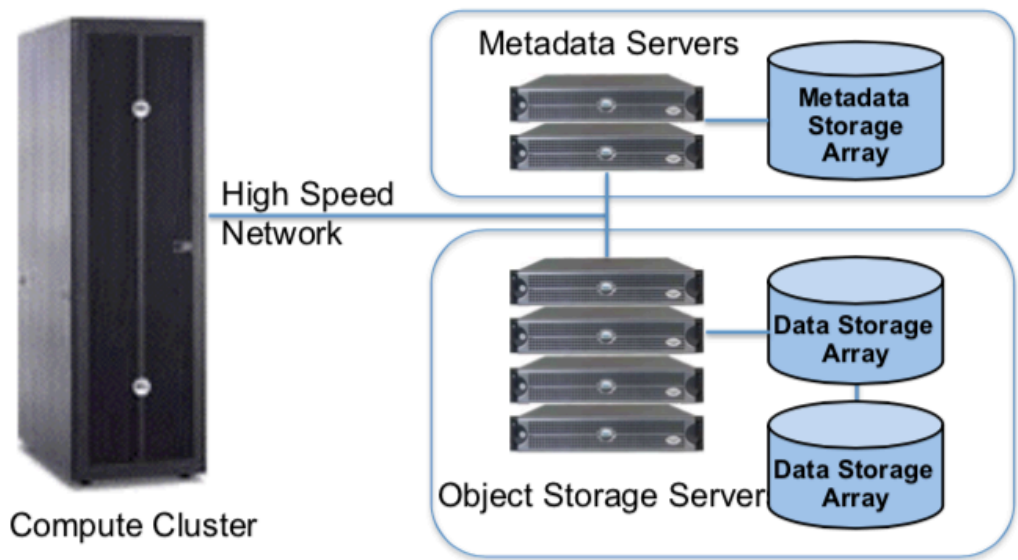

*Figure 3: A Lustre installation consists of metadata servers and object storage servers. These are connected to a compute cluster via a high speed interconnect such at 10Gb Ethernet or Infiniband.*

A typical Lustre installation might have many OSSes. In turn, each OSS can have a large number of drives that are often formatted in a RAID configuration (often RAID6) to allow for the failure of any two drives in an OSS. The many drives in an OSS allows data to be read in parallel at high bandwidth. File objects are striped across multiple OSSes to further increase parallel performance. The above redundancy is designed to give Lustre high availability while avoiding a single point of failure. Data loss can only occur if three drives fail in the same OSS prior to any one of the failures being corrected. For Lustre, the typical storage penalty to provide this redundancy is approximately 35%. Thus, a system with 6 petabytes of raw storage will provide 4 petabytes of data capacity to its users.

Lustre is designed to deliver high read and write performance to many simultaneous large files. Lustre systems offer very high bandwidth access to data. For a typical Lustre configuration, this bandwidth may be in excess of 100 of GB/second. This is achieved by the clients having a direct connection to the OSSes via a well-designed high-speed network. This connection is brokered by the MDS. The peak bandwidth of Lustre is determined by the aggregate network bandwidth to the client systems, the bisection bandwidth of the network switch, the aggregate network connection to the OSSes, and the aggregate bandwidth of the all the disks. Like most file systems, Lustre is designed for sequential read access and not random lookups of data (unlike a database). To find a particular data value in Lustre requires, on average, scanning through half the file system.

# 4. DATABASE MANAGEMENT SYSTEMS

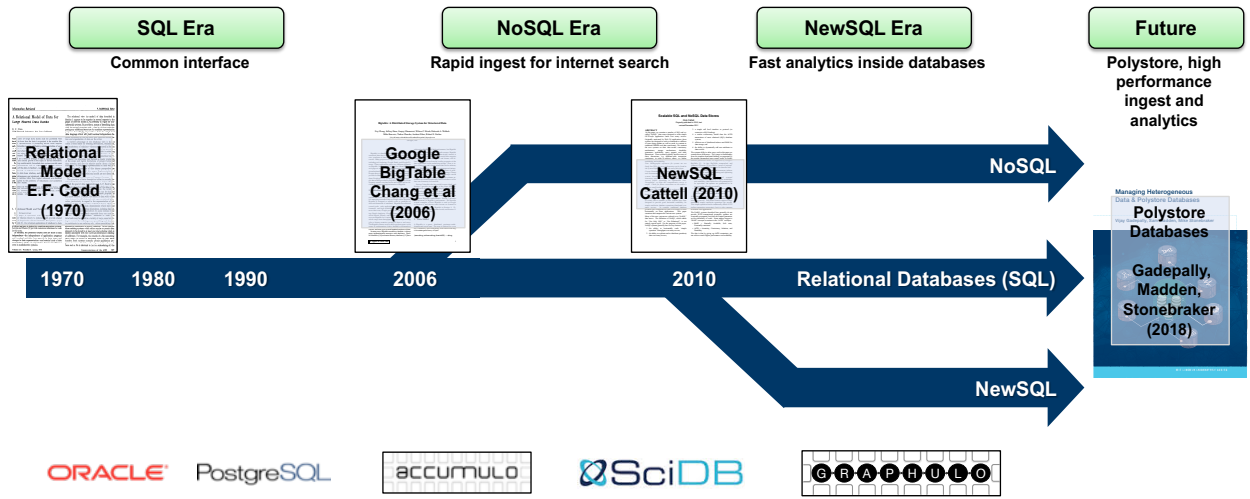

*Figure 4: High level view of database evolution*

Traditionally, database systems are seen as the natural data management approach. A database is a collection of data and supporting data structure. Traditionally, databases are exposed to users via a database management system. Users interact with these database management systems to define new data structures, schemas (data organization), to update data, and retrieve data. Beyond databases, developers may store data as files leveraging parallel file systems such as Lustre [32]. For the remainder of this section, however, we will focus on database systems such as those shown in the figure above. A more detailed view of data management systems is available in [8].

Traditional database management systems such as Oracle [34] and PostGRES [35], sometimes referred to as relational databases, while compliant with ACID [36] guarantees, are unable to scale horizontally for certain applications [37]. To address these challenges, large internet companies such as Google and Facebook developed horizontally scalable database technologies such as BigTable [38] and Cassandra [39]. These NoSQL [40] (not-only structured query language [SQL]) technologies enabled rapid ingest and high performance even on relatively modest computing equipment. BigTable inspired databases such as Apache Accumulo [41] extended the NoSQL model for application specific requirements such as cell-level security. NoSQL databases do not provide the same level of guarantees on the data as relational databases [37]; however, they have been very popular due to their scalability, flexible data model, and tolerance to hardware failure. In the recent few years, spurred by inexpensive high performance hardware and custom hardware solutions, we have seen the evolution of a new era in database technologies, sometimes called NewSQL databases [42]. These data management systems largely support the scalability of NoSQL databases while preserving the data guarantees of SQL-era database systems. Largely, this is done by simplifying data models, such as in SciDB, or leveraging in-memory solutions such as in MemSQL and Spark [43]. Looking towards the future, we see the development of new data management technologies that leverage the relative advantages of technologies developed within the various eras of database management technologies. A very high-level view of this evolution is presented in Figure 4. Looking towards the future, it is clear that no single type of database management systems is likely to support the kinds of data being collected from heterogenous sources of structured and unstructured data [44, 45]. Understanding how these different systems fundamentally interact with each other has a number of practical and theoretical [46], [47] implications. In order to address this challenge, one example of an active area of research in data management is in multi-database systems [48] such as Polystore databases [49] and a specific example is the BigDAWG system described in the next section.

Relational or SQL (Structured Query Language) databases [50] [51] have been the de facto interface to databases since the 1980s and are the bedrock of electronic transactions around the world. For example, most financial transactions in the world make use of technologies such as Oracle or dBase. With the great rise in quantity of unstructured data and analytics based on the statistical properties of datasets, NoSQL (Not Only SQL) database stores such as the Google BigTable [38] have been developed. These databases are capable of processing the large heterogenous data collected from the Internet and other sensor platforms. One style of NoSQL databases which have become used for applications that require support for high velocity data ingest and relatively simple cell-level queries are key-value stores.

As a result, the majority of the volume of data on the Internet is now analyzed using key-value stores such as Amazon Dynamo [52] and HBase [53]. Key-value stores and other NoSQL databases compromise on data consistency in order to provide higher performance. In response to this challenge, the relational database community has developed a new class of relational databases (often referred to as NewSQL [54]) such as SciDB [55], H-Store [56], VoltDB [57] to provide the features of relational databases while also scaling to very large data sets. Very often, these NewSQL databases make use of a different data model or advances in hardware architectures [54]. For example, MemSQL [58] is a distributed in-memory database that provides high performance, ACID compliant relational database management. Another example, BlueDBM [59], provides high performance data access through flash storage and field programmable gate arrays (FPGA).

## Database Management Systems and Features

A database is a collection of data and all of the supporting data structures. The software interface between users and a database is known as the database management system. Database management systems provide the most visible view into a dataset. There are many popular database management systems such as MySQL [60], PostgreSQL [35], and Oracle. Most commonly, users interact with database management systems for a variety of reasons:

- To define data, schema, and ontologies
- To update/modify data in the database
- To retrieve or query data
- To perform database administration or modify parameters such as security settings
- More recently, to perform analytics on the data within the database

Databases are used to support data collection, indexing and retrieval through transactions. A database transaction refers to the collection of steps involved in performing a single task [61]. For example, a single financial transaction such as "credit $100 towards the account of John Doe" may

involve a series of steps such as locating the account information for John Doe, determining the current account value, adding $100 to the account, and ensuring that this new value is seen by any other transaction in the future. Different databases provide different guarantees on what happens during a transaction.

Relational databases provide ACID guarantees: atomicity, consistency, isolation and durability. Atomicity provides the guarantee that database transactions either occur fully or completely fail. This property is useful to ensure that parts of a transaction do not occur successfully if other parts fail, which may lead to an unknown state. The second guarantee, consistency, is important to ensure that all parts of the database see the same data. This guarantee is important to ensure that when different clients perform transactions and query the database, they see the same results. For example, in a financial transaction, a bank account may be debited before further transactions can occur. Without consistency, parts of the database may see different amounts of money (not a great database property!). Isolation in a database refers to a mechanism of concurrency control in a database. In many databases, there may be numerous transactions occurring at the same time. Isolation ensures that these transactions are isolated from other concurrent transactions. Finally, database durability is the property that when a transaction has completed, it is persisted even if the database has a system failure. Non-relational databases such as NoSQL databases often provide a relaxed version of ACID guarantees referred to as BASE guarantees in order to support a distributed architecture or performance. This stands for Basically Available, Soft State, Eventual Consistency guarantees [62]. As opposed to the ACID guarantees of relational databases, non-relational databases do not necessarily provide strict guarantees on the consistency of each transaction but instead provide a looser guarantee that *eventually* one will have consistency in the database. For many applications, this may be an acceptable guarantee.

For these reasons, financial transactions employ relational databases that have the strong ACID guarantees on transactions. More recent trends that make use of the vast quantity of data retrieval from the Internet can be done via non-relational databases such as Google BigTable [38] which are responsible for fast access to information. For instance, calculating statistics on large datasets are not as susceptible to small eventual changes to the data.

While many aspects of learning how to use a database can be taught through books or guides such as this, there is an artistic aspect to their usage as well. More practice and experience with databases will help overcome common issues, improved performance tuning, and help with improved database management system stability. Prior to using a database, it is important to understand the choices available, properties of the data and key requirements.

Based on a series of papers from Google in the mid-2000s, the MapReduce computing paradigm was created which gained wide acceptance through the open source Apache Hadoop soon after. These technologies, combined with the seminal Google BigTable [38] paper helped spark the NoSQL movement in databases. Not long after this, numerous technologies such as GraphLab [63], Neo4j [64], and Giraph [65] were developed to apply parallel processing to large unstructured graphs such as those being collected and stored in NewSQL databases. Since the year 2010, there has been renewed interest in developing technologies that offer high performance along with some of the ACID guarantees of relational databases. This requirement has driven the development of a new generation of relational databases often called NewSQL. In the parallel processing world, users are looking for better ways to deal with streaming data or machine learning and graph algorithms than the Hadoop framework offered and are developing new technologies such as Apache Storm [66], Spark [67], and Graphulo [68], [69]. A more detailed view of these techniques is available in [29].

**Relational Databases**

Relational databases such as MySQL, PostgreSQL, and Oracle form the bedrock of database technologies today. They are by far the most widely used and accessed databases. We interact with these databases daily: everywhere financial transactions, medical records, and purchases are made. From the CAP theorem [70], which does have known issues [71, 72], relational databases provide strong consistency and availability; however, they do not support partition tolerance. In order to avoid issues with partition tolerance in distributed databases, relational databases are often vertically scalable. Vertical scalability refers to systems that scale by improving existing software or hardware. For example, vertically scaling a relational database involved improving the resources of a single node (more memory, faster processor, faster disk drive, etc.). Thus, relational databases often run on high-end, expensive nodes and are often limited by the resources of a single node. This is in contrast to non-relational database that are designed to support horizontal scalability. Scaling a database horizontally involves adding more nodes to the system. Most often, these nodes can be inexpensive commercial off-the-shelf systems (COTS) that are easy to add as resource requirements change.

Relational databases provide ACID guarantees and are used extensively in practice. Relational databases are called *relational* because of the underlying data model. A relational database is a collection of tables that are connected to each other via relations expressed as keys. The specification of tables and relations in a database is referred to as the schema. Schema design requires thorough knowledge of the dataset. Many databases may contain tens to hundreds of tables and require careful thought during the design.

**NoSQL Databases**

Since the mid-2000s and the Google BigTable paper, there has been a rise in popularity of Not Only SQL (NoSQL) databases. NoSQL databases support many of the large scale computing activities with which we interact regularly such as web searches, document indexing, large-scale machine learning, and graph algorithms. NoSQL databases support horizontal scaling: you can increase the performance through the addition of nodes. This allows for scaling through the addition of inexpensive commercial off-the-shelf systems as opposed to expensive hardware upgrades required for vertical scaling. NoSQL databases often need to relax some of the consistency or availability guarantees of relational databases in order to take advantage of strong partition tolerance guarantees. In order to keep up with rising data volumes, organizations such as Google looked for ways to incorporate inexpensive off-the-shelf systems for scaling their hardware. However, incorporating such systems requires the use of networks which can be unreliable. Thus, partition tolerance to network disruptions became an important design criterion. In keeping with the CAP theorem, either consistency or availability must be relaxed to provide partition tolerance in a distributed database (though this is likely an oversimplification [73]).

At a transaction level, NoSQL databases may provide BASE guarantees. These guarantees may not be suitable for many applications where strong consistency or availability is required. Of course, before choosing a technology to use for an application, it is important to be aware of all design constraints and the impact of technology choice on the final analytic requirements.

**NewSQL Databases**

The most recent trend in database design is often referred to as NewSQL databases. Given the controversy surrounding the CAP theorem, such databases attempt to provide a version of all three distributed database properties. These databases were created to approach the performance of NoSQL databases while providing the ACID transaction guarantees of traditional relational databases [58]. In order to provide this combination, NewSQL databases often employ different hardware or data models than traditional database management systems. NewSQL databases may be considered as an alternative to both SQL and NoSQL style databases [43]. Most NewSQL databases provide support for the Structured Query Language (SQL) or other popular query languages.

## Access Control

Access control in databases has played an integral role in their development and popularity [74] [75]. Access control can be used at granularities such as the level of a table, individual rows, or even individual cells. Depending on the access control implementation, controlling which users can access what data entries can be a challenging task [76] sometimes amplified by application-specific requirements [77]. In general, database access control limits the access of a principal, a user or users, to the contents of a database. We separate the concepts of access control strategies and access control mechanisms.

An access control refers to how access control policies are assigned to principals, whereas an access control *mechanism* determines how access to the database is restricted. View-based access control, the most common access control mechanism found in production database systems, is data-dependent and is often implemented through metadata.

For example, role-based strategies can be applied to view-based or query-based mechanisms. For the most part, relational database management systems have concentrated on view-based mechanisms with varying access control strategies. Therefore, these terms are often conflated and role-based access control in many contexts implies a view-based access control mechanism with a role-based strategy. It should be noted that multiple access control mechanisms need not be mutually exclusive. On a single database, query control can be used to limit the queries that are actually executed, and view-based access control can be used to limit the results returned.

In a view-based database access control model, a principal requests access to database contents. The system evaluates whether the principal is authorized to access the database contents by examining the access control policy. Often, an access control policy depends on the contents being accessed. The system issues a decision that either allows or denies access. View-based access control uses a database view as an abstraction mechanism for the data available to a particular principal [75]. There are a number of historical models for access control strategies applied to the view-based access control mechanism and some early strategies, such as Discretionary Access Control and Mandatory Access Control [78] were often implemented via individual or group level access control. Role-based access control is a popular way to implement access control policies [79]. Recently, there has been research in a new paradigm for access control – query control [80]. Query control places a restriction on the queries that a principal can issue, and is therefore not data-dependent. Access control strategies are orthogonal to access control mechanisms. A more detailed view of access control trends is available in [80].

Additionally, one may wish to provide layered security on top of the database to protect the confidentiality, integrity and/or availability of data. For examples, tools such as CryptDB [81] and Computing on Masked Data [9, 13] are tools that support storing and retrieving encrypted data in SQL and NoSQL databases, respectively. An overview of security within the context of database systems is given in [82].

## 5. HETEROGENOUS DATA MANAGEMENT

Modern applications often need to manage and analyze widely diverse datasets that span multiple data models. In medical informatics [83] [84], health professionals serve patients admitted to intensive-care units using data expressed as structured demographics, semi-structured laboratory and microbiology test results, discharge summaries, radiology, cardiology reports in text formats, and vital signs and other data in time-series format. In oceanographic metagenomics [85], biologists detect relationships between cyanobacteria communities and environmental parameters via integrating genome sequences, structured sensor and sample metadata, cruise reports in text formats, and streaming data generated by flow-cytometer systems. In intelligent transportation management [86] administrators analyze open traffic data presented in RDF (Resource Description Framework), city events expressed as JSON (JavaScript Object Notation) documents, social-media data recorded via key-value pairs, and weather feeds stored in relational tuples to predict traffic flows. Finally, in data journalism [87], journalists work with Tweet texts, relational databases provided by governments and institutions, and RDF-formatted Linked Open Data to support content management for writing political articles.

In these and other scenarios, warehousing the data using Extract-Transform-Load (ETL) processes can be very expensive. First, transforming disparate data into a single chosen data model may degrade performance. Indeed, there appears to be no ``one size fits all'' solution for all markets [45] [44] as specialized models and architectures enjoy overwhelming advantages in data warehousing, text searching, stream processing, and scientific databases. Second, curating diverse datasets and maintaining the pipeline could turn out to be labor intensive [88]. One major reason is that rules and functions in ETL scripts do not adapt to changes in data and analytical requirements, and changes in application logic often result in the modification of ETL scripts.

For these and other reasons, a number of projects are shifting the focus to federating specialized data stores and enabling query processing across heterogeneous data models [49]. This shift can bring many advantages. First, the systems can build natively on multiple data models, which can translate to maximizing the semantic expressiveness of underlying interfaces and to leveraging the internal processing capabilities of each data store. Typical tasks can be expressed natively in a variety of algebras, such as relational, linear, and graph algebra, and be executed economically on a variety of specialized data stores optimized for different workloads. Second, federated architectures support query-specific data integration with just-in-time transformation and migration, which has the potential to significantly reduce the operational complexity and overhead. Data transformations across data models and data migration between data stores can be explicitly expressed via queries and automatically handled by the system, bridging the gap between data preparation and data analysis.

Projects that focus on developing systems in this research area stem from various backgrounds and address diverse concerns, which can make it difficult to form a consistent view of the work in the area. Some of the projects concentrate on the issues of semantic mapping and record linkage; some define operators over multiple data models and focus on multi-model query planning and optimization; others emphasize data-flow optimization and multi-platform scheduling. Such diverse perspectives and viewpoints add to the complexity of understanding the field, and might even cause unnecessary miscommunication between research groups. Therefore, it would be beneficial to have a taxonomy of the field that would contribute to clear definitions of the key terms. We could then build on the taxonomy by specifying an evaluation framework focused on the query-processing characteristics of each design.

Systems federating specialized data stores and enabling query processing across heterogeneous data models can be characterized by the data stores and query interfaces that they support. We introduce a taxonomy that builds on this observation and groups state-of-the-art solutions into four categories, defined as follows:

- A federated database system comprises a collection of homogeneous data stores and features a single standard query interface. Example: Multibase [89].
- A polyglot system hosts data using a collection of homogeneous data stores and exposes multiple query interfaces to the users. Example: Spark SQL [90]
- A multistore system is able to manage data across heterogeneous data stores, while supporting a single query interface. Example: Polybase [91], D4M [41, 92]
- A polystore system enables query processing across heterogeneous data stores and supports multiple query interfaces. Example: BigDAWG [37]

A deeper view into these trends and heterogenous data management systems is available in [48].

## Polystore System Deep Dive

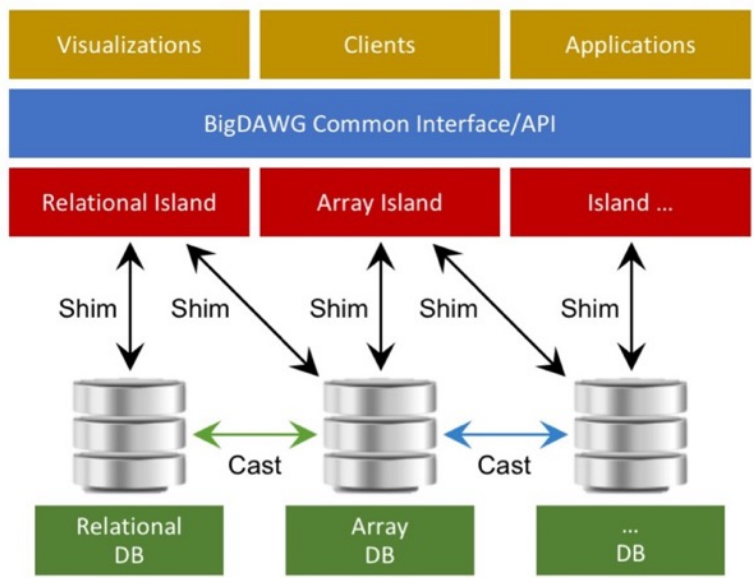

*Figure 5: The BigDAWG architecture*

BigDAWG [37, 85, 93], short for the Big Data Working Group, is an implementation of a polystore database system designed to simplify database management for complex applications. For example, modern decision support systems are required to integrate and synthesize a rapidly expanding collection of real-time data feeds: sensor data, analyst reports, social media, chat, documents, manifests, logistical data, and system logs (to name just a few). The traditional technique for solving a complex data fusion problem is to pick a single general-purpose database engine and move everything into this system. However, custom database engines for sensors, graphs, documents, and transactions (just to name a few) provide 100× better performance than general-purpose databases. The performance benefits of custom databases have resulted in the proliferation of data-specific databases, with most modern decision support systems containing five or more distinct customized storage systems. Additionally, for organizational or policy reasons, data may be required to stay in disparate database engines. For an application developer, this situation translates to developing his or her own interfaces and connectors for every different system. In general, for N different systems, a user will have to create nearly $N^2$ different connectors. BigDAWG allows users to access data stored across multiple databases via a uniform common interface. Thus, for a complex applications in which there is scientific data, text data, and metadata, a user can store each of these components in the storage technology best suited to each data type, but also develop analytics and applications that make use of all of these data without having to write custom connectors to each of these storage technologies. The end-to-end architecture of the BigDAWG polystore system is described in Figure 5. This architecture describes how applications, visualizations, and clients at the top access information stored in a variety of database engines at the bottom. At the bottom, we have a collection of disparate storage engines (we make no assumption about the data model, programming model, etc., of each of these engines). These storage engines are organized into a number of islands. An island is composed of a data model, a set of operations, and a set of candidate storage engines. An island provides location independence among its associated storage engines. A shim connects an island to one or more storage engines. The shim is basically a translator that maps queries expressed in terms of the operations defined by an island into the native query language of a particular storage engine. A key goal of a polystore system is for the processing to occur on the storage engine best suited to the features of the data. We expect in typical workloads that queries will produce results best suited to particular storage engines. Hence, BigDAWG needs a capability to move data directly between storage engines. We do this with software components we call casts.

**Database and Storage Engines**

A key design feature of BigDAWG is the support of multiple database and storage engines. With the rapid increase in heterogeneous data and the proliferation of highly specialized, tuned, and hardware-accelerated database engines, it is important that BigDAWG support as many data models as possible. Further, many organizations already rely on legacy systems as a part of their overall solution. We believe that analytics of the future will depend on many distinct data sources that can be efficiently stored and processed only in disparate systems. BigDAWG is designed to address this need by leveraging many vertically integrated data management systems. The current implementation of BigDAWG supports a number of popular database engines: PostGRES (SQL), MySQL (SQL), Vertica (SQL), Accumulo (NoSQL), SciDB (NewSQL), and S-Store (NewSQL). The modular design allows users to continue to integrate new engines as needed.

### BigDAWG Islands

The next layer of the BigDAWG stack is its islands. Islands allow users to trade off between semantic completeness (using the full power of an underlying database engine) and location transparency (the ability to access data without knowledge of the underlying engine). Each island has a data model, a query language or set of operators, and one or more database engines for executing them. In the BigDAWG prototype, users determine the scope of their query by specifying an island within which the query will be executed. Islands are a user-facing abstraction, and they are designed to reduce the challenges associated with incorporating a new database engine. The current implementation of BigDAWG supports islands with relational, array, text, and streaming models. Our modular design supports the creation of new islands that encapsulate different programming and data models.

### BigDAWG Middleware and API

The BigDAWG "secret sauce" lies in the middleware that is responsible for developing cross-engine query plans, monitoring previous queries and performance, migrating data across database engines as needed, and physically executing the requested query or analytic. The BigDAWG interface provides an API to execute polystore queries. The API layer consists of server- and client-facing components. The server components incorporate islands that connect to database engines via lightweight connectors referred to as shims. Shims essentially act as an adapter to go from the language of an island to the native language of an underlying database engine. In order to identify how a user is interacting with an island, a user specifies a scope in the query. A scope of a query allows an island to correctly interpret the syntax of the query and allows the island to select the correct shim that is needed to execute a part of the query. Thus, a cross-island query may involve multiple scope operations. Details of the BigDAWG middleware can be found in [94]–[95] [96] [97].

# 6. DEVELOPING A WORKING DATA HUB

This section outlines some of the key considerations when developing a working data hub. Some of the key ingredients include: Technology, Infrastructure, Formatting, Security, Policy, and User Outreach.

## Technology Considerations

By definition, a data hub is comprised of multiple heterogenous computing and storage units distributed across a local or wide area network. Many of the traditional approaches to implementing a data hub relied on a centralized data lake. As time has gone one, these data lake solutions have had a number of issues that have made them unpopular with data owners and users:

- Single point of failure: If the system goes down for scheduled or unscheduled maintenance, all users who rely on the system are affected. This also makes it difficult to use the data lake for operational data collection and data is often collected in a different system.
- Limited data reuse: Once data is pushed to the data lake, it is often difficult to update and data often becomes stale.
- Expense: Having a centralized solution (data lake) can be expensive in terms of cost, maintenance and space

### Technology Recommendations

Given these considerations, we recommend organizations invest in data hubs that:

- Support federated data access: Leverage a scalable data management architecture that allows the addition of new technologies such as object stores, databases and file systems. It is unlikely that any single technology solution will scale or provide efficient access to all modalities of data and federation as an architectural principle is important.
- Support Complex access control: Most organizations will have complex access control. For example, certain users may be allowed to issue certain queries but only if the query produces a small number of results. Further, given mission requirements, access control may change on a daily basis. New techniques such as query-based access control can be used to mitigate such challenges.
- Tie compute with data: Provide seamless access to the computing infrastructure that will be used to process the data; train machine learning models, etc.
- Support Multiple data formats: Different data types, sensors and data owners will leverage different data formats. Technology that supports a data hub should support these diverse data formats. For example, potential technology should support video, imagery, time-series and text data. While some of these modalities are not covered in a data catalog, it is important to highlight these differences up front.
- ML models should not be forgotten: Storing and providing a uniform access mechanism to data sets is an important first step. Additionally, developers should keep in mind that machine learning models also need a place to be stored. Often, the cost in developing these ML models can be similar to data collection. (See Appendix: X)

- A designed with technical debt awareness [98]: Using unproven or unsupported technology can lead to long-term challenges in maintaining a working system. Be aware of technology developers and dependencies required to run software. Additionally, technical debt can be amplified in machine learning and artificial intelligence applications [99].

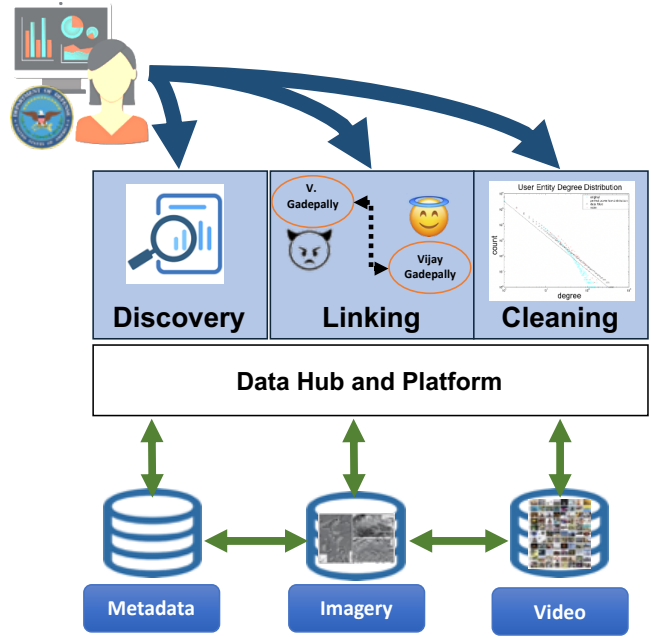

*Figure 6: Data hub components*

## Infrastructure Considerations

A critical component of providing seamless services is the layering of robust infrastructure. This infrastructure includes computing nodes, storage servers and networking equipment. Even the best tools can often be limited by under provisioned hardware.

### Infrastructure Recommendations

- Ensure that you have sufficient infrastructure and computing capabilities: Modern cloud environments often consist of PB of data storage and hundreds of computing nodes. An under provisioned system cannot scale with the rate of new data.
- Cloud computing is a potential infrastructure solution, but be aware of long-term costs and cloud hooks: Many organizations are currently in the process of "cloud clawback" and leveraging hybrid public and private cloud options. When evaluating a cloud solution, keep in mind that certain technical decisions may lock you into a single vendor. For example, using software or hardware solutions that are only available in a single cloud provider.
- Be prepared to run multiple versions of the same tools on different networks: Sensitivity of of data can add additional complexity. It is desirable to have the same software tools available on sensitive as well as open networks. Pay particular attention to software and hardware tools. For example, one may wish to use a particular database backend that is not supported or approved for use on a sensitive network.
- To maximize the utility of data in the Data Hub, it's important to have a data catalog. Such a catalog would be updated automatically (perhaps by the data discovery mechanism) and provide search capability based on metatags associated with the data. The catalog could additionally link to models or analysis / processing code that has been used successfully.   Such a catalog enables the user to build on what others have done in a way that more collecting data in one place cannot.

## Data Formatting Considerations

There are simple techniques that can be applied during initial parsing of raw data that can dramatically reduce the effort of applying AI.  This parsing is much more efficient to do during initial collection setup when the knowledge of the data exists with the programmer.   Requiring an AI analyst to later deduce this knowledge is a primary reason why "data wrangling" is often 80% of the effort in building an AI system.

Key challenges:
- 80% of researcher time developing machine learning and artificial intelligence solutions spent on data plumbing, wrangling, archaeology, etc.
- Often original data creators are unavailable and important data meaning is lost

### Formatting Recommendations
- Data creators should store data in human and machine understandable tabular formats
- XML (viewable)
- JSON (viewable, readable)
- Tables* (viewable, readable, and <u>understandable</u>)
- Human understandable formats are valuable for ISO sign off
- Stick with simple and uniform conventions for file naming
- Allows spot checking of products and removal of columns

## Security/Data Sharing Considerations

Data is a valued commodity. Having the trust of end-users, data providers and security is a critical element to a successful data hub. A few challenges:

- Sharing AI Data often requires Information Security Officer (ISO) sign-off
- ISOs and Subject Matter Experts (SMEs) have different terminology
- ISOs sign off requires confidence in SME data handling practices
- ISOs need basic information to allow data sharing: project, need, location, personnel, duration, ...
- SMEs often provide research descriptions that limit ISO security surety in SMEs data handling practices and results in ISOs limiting of data sharing requests

### Security/Sharing Recommendations
- Work with ISOs from the beginning: Keep ISO abreast of data sharing plans from the beginning. Often, very minor changes made early on can improve likelihood of eventual ISO signoff.
- Integrate SME feedback with ISO requirements

- Data Aggregation: Aggregating data may lead to security issues. For example, multiple non sensitive sources, when aggregated, may lead to sensitive data. For example, in the health domain, combining information about dates of birth and names can lead to personally identifiable information. Providers of data aggregation services should be aware and provide mitigations. For example, data aggregation may need to occur only on approved systems allowed to process protected data.
- Provide support for data use agreements when sharing with external parties: In the academic/commercial world, data use agreements can provide a mechanism through which data owners, SMEs, and ISOs can be aware of who is using the data, the purpose of using the data, and the duration of data use. Such an agreement ensures that only legitimate users of data have access to the data. A template agreement is shown in Figure 7.

## Policy Considerations:

These are considerations for decision makers in formulating policy that can speed-up data sharing and success of a data hub. These policies often have to do with incorporating new technologies and enabling developers to keep up with the pace of new demands.

### Policy Recommendations

- Acquisition environment: As new data is collected it will be necessary to integrate new technologies. This rate of change may be faster than traditional processes are able to support.
- Prioritize open-source vs. closed-source software products: However, be aware of availability of long-term support and a robust developer community. The Navy may also consider supporting key personnel of open-source software to ensure feature availability and development.
- Have Top-Down technology selection. Involve management in technology selection process. Avoid products/technologies that have unknown/unreliable development team (e.g., teams for adversary nations; teams of individuals unlikely to continue maintaining software)
- Be aware of software licensing: Certain software libraries and products have restrictive software licenses. This may limit the ability to share technology with other industry/academic partners.
- Avoid software/hardware products with unknown user base or non-active developer base. Reevaluate software/hardware products when the developers are acquired by other entities.

## User Engagement and Incentives

In order to prevent a data hub from becoming stale and unused, it is imperative to have an active and engaged base of users and developers. For this, it is important to first build an ecosystem that leverages stakeholder feedback and provides timely additions to satisfy a diverse user base.

**User Engagement Recommendations**

- Incorporate user feedback early in the development process to survey needs and wants
- Make data sharing and maintenance a key part of promotions/performance reviews across the enterprise. For example, a performance bonus for data sharing comparable to those received for publications and patents.
- Insist on data sharing as a part of large-scale projects. For example, if funding new initiatives, ask what the data sharing plan will be up front.
- Remove cost from the data sharing equation. Avoid chargeback models in which programs are responsible for upkeep and maintenance costs associated with data sharing.
- Leverage open non-proprietary standard
- Involve key influencers across the organization along with incentives for users and developers
- Invest in data discovery techniques. Most current cataloging systems require users to actively push their data and maintain it. A data discovery system could greatly automate such processes. Such a data discovery system could include crawlers that go through a network to look for interesting datasets. By using associated metadata, this discovery system may be able to pull together rudimentary information that could be used to automate catalog creation. An example of this at the internet scale is given in [23].

**Sample Data Use Agreement**

*When communicating a data release request with an information security officer (ISO), the following topics should be kept in mind and touched upon in the initial communication. While it is good to have additional information available if follow-ups are requested, the initial communication should be kept fairly short and minimize the use of domain specific terminology.*

What is the data you're seeking to share?
Describe the data to be shared, focusing on its risk to the organization if it were to be accidently released to the public or otherwise misused.
Example:
> *The data was collected on <<date range>> at <<location(s)>> in accordance with our mission. The risk has been assessed and addressed by an appropriate combination of excision, anonymization, and/or agreements. The release to appropriate legitimate researchers will further our mission and is endorsed by leadership.*
> *Explanation:*
> *Sentence 1 establishes the identity, finite scope, and proper collection of the data. Sentence 2 establishes that risk was assessed and that mitigations were taken. Sentence 3 establishes the finite scope of the recipients, an appropriate reason for release, and mission approval.*

Where / to whom is the data going?
Please describe the intended recipients of the data, the systems they will use to receive / process the data.
Example:
> *The data will be shared with researchers at <<institution>>. The data will be processed on <<institution>> owned systems meeting their institution security policies, which include password controlled access, regular application of system updates, and encryption of mobile devices such as laptops. Authorized access to the data will be limited to personnel working as part of this effort.*
> *Explanation:*
> *Sentence 1 establishes the legal entity trusted with the data and with whom any agreements are ultimately made on behalf of. Sentence 2 establishes that basic technical safeguards are in place, without getting too specific, and that personally-owned computers will not be used as the institution has no legal control over them. Sentence 3 establishes that the data will not be used for other purposes than the agreed-upon research project.*

What controls are there on further release (policy/legal & technical)?

Is a non-disclosure or data usage agreement in place?

Is the data anonymized? If so, is there an agreement in place to prohibit de-anonymization attempts?

What technical controls are in place on the systems that will receive / process the data to prevent misuse?

Is there an agreement in place on publication of results from this effort?

Is there an agreement in place for the retention or deletion of the original data, intermediate products, and/or the results at the end of the effort?

Example:
> *An acceptable use guideline that prohibit attempting to de-anonymize the data and will be provided to all personnel working on the data. Publication guidelines have been agreed to that allow for high-level statistical findings to be published, but prohibit including any individual data records. A set of notional records has been provided that can be published as an example of the data format, but is not part of the actual data set. The research agreement requires all data to be deleted at the end of the engagement except those items retained for publication.*
> *Explanation:*
> *Sentence 1 establishes that there is an agreement in place prohibiting de-anonymizing the data and clearly defining it as "misuse" of the data. Sentence 2 and 3 establish that it is known to all parties what may and may not be published. Sentence 4 establishes that data retention beyond the term of the agreement has been addressed and cleanup is planned as part of project closeout.*

*Figure 7: Template Data Use Agreement*